\documentclass{article}
\usepackage{spconf,amsmath,graphicx}
\usepackage{dsfont}
\usepackage{multirow}
\usepackage{arydshln}
\usepackage{booktabs}
\usepackage{hyperref}
\usepackage{enumitem}
\usepackage[tracking=true]{microtype}
\usepackage[table,xcdraw]{xcolor}
\usepackage{boldline}
\usepackage{enumitem}
\usepackage{setspace}
\usepackage{slantsc}
\usepackage[export]{adjustbox}
\usepackage[lofdepth,lotdepth]{subfig}
\usepackage[tracking=true]{microtype}

\title{CDPAM: Contrastive learning for perceptual audio similarity}

\name{Pranay Manocha$^{1}$, \hspace{0mm} Zeyu Jin$^{2}$, \hspace{0mm} Richard Zhang$^{2}$, \hspace{0mm} Adam Finkelstein$^{1}$}

\address{$^{1}$Princeton University, USA \hspace{30mm}$^{2}$Adobe Research, USA}
\newcommand{\ignorethis } [1] {}

\newcommand{\etal       }     {{et~al.}}

\newcommand{\eg         }     {{e.g.}}

\newcommand{\Reals      }     {{\textrm{I\kern-0.18em R}}}

\newcommand{\change     } [1] {\mbox{{\footnotesize $\Delta$} \kern-3pt}#1}

\newlength{\w}

\newcommand\narrowstyle{\SetTracking{encoding=*}{-50}\lsstyle}
\newcommand\normalstyle{\SetTracking{encoding=*}{0}\lsstyle}

\newcommand{\acronym}[1] {\narrowstyle{\scshape{#1}}\normalstyle}

\newcommand{\PESQ}   {\acronym{Pesq}}
\newcommand{\VISQOL} {\acronym{Visqol}}

\newcommand{\SESQA}  {\acronym{Sesqa}}
\newcommand{\DPAM}   {\acronym{Dpam}}
\newcommand{\OURS}   {\acronym{Cdpam}}

\begin{document}
\ninept
\maketitle
\begin{abstract}

Many speech processing methods based on deep learning require an automatic and differentiable audio metric for the loss function. 
The \DPAM\ approach of Manocha~\etal~\cite{manocha2020differentiable} learns a full-reference metric trained directly on human judgments, and thus correlates well with human perception. 
However, it requires a large number of human annotations and does not generalize well outside the range of perturbations on which it was trained.
This paper introduces \OURS~-- a metric that builds on and advances \DPAM. 
The primary improvement is to combine contrastive learning and multi-dimensional representations to build robust models from limited data. 
In addition, we collect human judgments on triplet comparisons to improve generalization to a broader range of audio perturbations. 
\OURS\ correlates well with human responses across nine varied datasets.
We also show that adding this metric to existing speech synthesis and enhancement methods yields significant improvement, as measured by objective and subjective tests.

\end{abstract}
\begin{keywords}
perceptual similarity, audio quality, deep metric, speech enhancement, speech synthesis
\end{keywords}
\section{Introduction}
\label{sec:intro}

% par 1: smoothing out the wording
% par 2: move some of the discussion to related work
%   DPAM - correlates well
%   limitations:
%     1. requires lots of expensive data
%     2. somewhat robust but generalization could be improved: 
%       (a) content variations (speaker, text) because the learned model (see end of 2.1)
%       (b) away from JND
%
%  this paper builds on the DPAM framework and dataset using two ideas: contrastive learning and MDRL
%  contrastive learning is a form of self-supervised learning that augments cases 

% The task of audio similarity is to measure similarity between audio segments. Humans are great at it, while machine judgment still remains far behind human judgment. Objective metrics that correlate well with human judgment open the possibility to scale up automatic quality assessment, with consistent results at a negligible fraction of effort, time and cost of their subjective counterparts.

Humans can easily compare and recognize differences between audio recordings, but automatic methods tend to focus on particular qualities and fail to generalize across the range of audio perception. 
Nevertheless, objective metrics are necessary for many automatic processing tasks that cannot incorporate a human in the loop. 
Traditional objective metrics like \PESQ\,\cite{rix2001perceptual} and \VISQOL\,\cite{hines2015visqol} are automatic and interpretable, as they
rely on complex handcrafted rule-based systems. 
Unfortunately, they can be unstable even for small perturbations -- and hence correlate poorly with human judgments.

Recent research has focused on learning speech quality from  
data~\cite{manocha2020differentiable,lo2019mosnet,patton2016automos,fu2019learning,fu2019metricgan,serra2020sesqa}. 
One approach trains the learner on subjective quality ratings, \eg, MOS scores on an absolute (1 to 5) scale~\cite{lo2019mosnet,patton2016automos}. 
Such ``no-reference'' metrics support many applications but do not apply in cases that require comparison to ground truth, for example for learned speech synthesis or enhancement~\cite{kumar2019melgan,defossez2020real}.

Manocha~\etal~\cite{manocha2020differentiable} propose a ``full-reference'' 
deep perceptual audio similarity metric (\DPAM) trained on human judgments.
As their dataset focuses on just-noticeable differences (JND), the learned model correlates well with human perception, even for small perturbations. 
However, the \DPAM\ model suffers from a natural tension between the cost of data acquisition and generalization beyond that data. It requires a large set of human judgments to span the space of perturbations in which it can robustly compare audio clips. Moreover, inasmuch as its dataset is limited, the metric may generalize poorly to unseen speakers or content. Finally, because the data is focused near JNDs, it is likely to be less robust to large audio differences.

To ameliorate these limitations, this paper introduces \OURS: a \underline{c}ontrastive learning-based multi-dimensional \underline{d}eep \underline{p}erceptual \underline{a}udio similarity \underline{m}etric.
The proposed metric builds on \DPAM\ using three key ideas: (1) contrastive learning, (2) multi-dimensional representation learning, and (3) triplet learning.
Contrastive learning is a form of self-supervised learning that augments a limited set of human annotations with synthetic (hence unlimited) data: perturbations to the data that should be perceived as different (or not).
We use multi-dimensional representation learning to separately model
\emph{content} similarity (\eg~among speakers or utterances) and
\emph{acoustic} similarity (\eg~among recording environments).
The combination of contrastive learning and multi-dimensional representation learning allows \OURS\ to better generalize across content differences (\eg~unseen speakers) with limited human annotation~\cite{van2019disentangled,higgins2018towards}.
Finally, to further improve robustness to large perturbations (well beyond JND), we collect a dataset of judgments based on triplet comparisons, asking subjects: ``Is A or B closer to reference C?''

We show that \OURS\ correlates better than \DPAM\ with MOS and triplet comparison tests across nine publicly available datasets.
We observe that \OURS\ is better able to capture differences due to subtle artifacts recognizable to humans but not by traditional metrics. 
Finally, we show that adding \OURS\ to the loss function yields improvements to the existing state of the art models for speech synthesis~\cite{kumar2019melgan} and enhancement~\cite{defossez2020real}.
The dataset, code, and resulting metric, as well as listening examples, are available here:
\\{\footnotesize\tt\narrowstyle \href{https://pixl.cs.princeton.edu/pubs/Manocha_2021_CCL/}{https://pixl.cs.princeton.edu/pubs/Manocha\_2021\_CCL/}}

%{https://percepaudio.cs.princeton.edu/Manocha\20\_CDPAM/}}

\iffalse
\begin{figure*}[h!]
\vspace{-1\baselineskip}
\centering
\setlength{\w}{0.6\columnwidth}
\setlength{\tabcolsep}{4pt}
\begin{tabular}{c c c}
\subfloat[][\bf Self-supervised audio encoder]{\includegraphics[width=0.41\textwidth]{pictures/fig1/pre-train.png}
\label{fig1:subfig1}} &

%\includegraphics[width=0.41\textwidth]{pictures/model_arch_1.png} &

\subfloat[][\bf JND Training]{\includegraphics[width=0.21\textwidth]{pictures/fig1/JND.png}\label{fig1:subfig2}}
 &

\subfloat[][\bf Triplet fine-tuning]{\includegraphics[width=0.20\textwidth]{pictures/fig1/fine-tune.png}\label{fig1:subfig3}}
 &
%
%\bf a) Self-supervised audio encoder & \bf b) JND training & \bf c) Triplet fine-tuning \\

\iffalse
\begin{minipage}{\w}
\vspace{2pt}
\begin{spacing}{0.9}
{\footnotesize a) fig 1}
\end{spacing}
\end{minipage} &
%
\begin{minipage}{\w}
\centering
{\footnotesize b) fig 2 } 
\end{minipage} &
%
\begin{minipage}{\w}
\vspace{2pt}
\begin{spacing}{0.9}
{\footnotesize  c) fig 3}
\end{spacing}
\end{minipage}
%
\fi

\end{tabular}
\vspace{-1ex}
\caption{\textbf{Metric Architecture}. We first train an audio encoder using contrastive learning, then train the loss-net on JND data~\cite{manocha2020differentiable} keeping the encoder weights fixed. Finally, we fine-tune the loss-net on the newly collected dataset of triplet comparisons.}
\label{model_architecture}
\vspace{-3ex}
\end{figure*}

\fi

\begin{figure*}[h!]
\vspace{-2\baselineskip}
\centering
\setlength{\tabcolsep}{4pt}
\includegraphics[width=\textwidth]{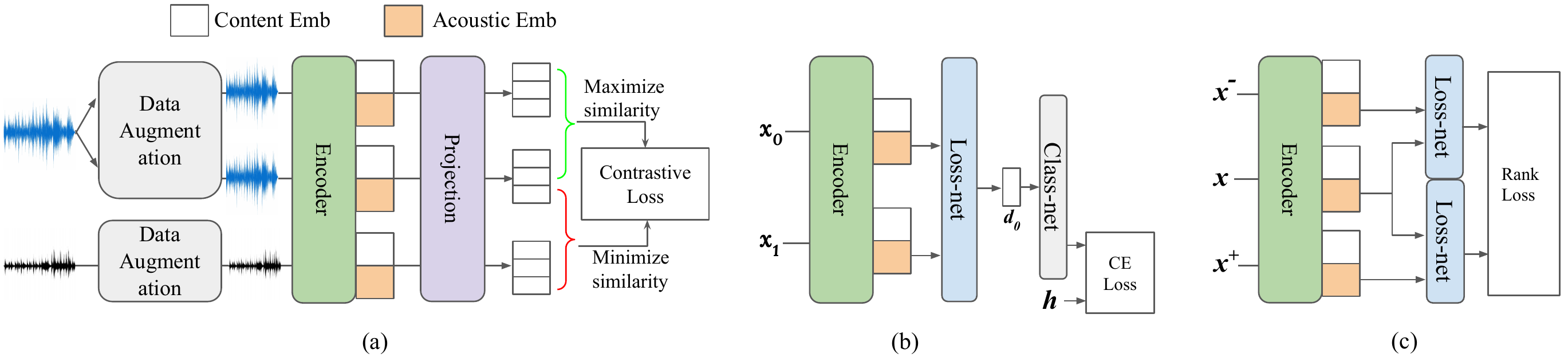}
\vspace{-0.25in}
\caption{\textbf{Training architecture}: (a)~We first train an audio encoder using contrastive learning, then (b)~train the loss-net on JND data~\cite{manocha2020differentiable}. (c)~Finally, we fine-tune the loss-net on the newly collected dataset of triplet comparisons.}
\vspace{-0.1in}
\label{model_architecture}
\vspace{-1ex}
\end{figure*}

\iffalse
\begin{figure*}[h!]
\vspace{-7\baselineskip}
\centering
\setlength{\tabcolsep}{4pt}
\includegraphics[width=\textwidth]{pictures/model_architecture.png}\\

%\bf \hspace{0.5cm} Self-supervised Audio encoder \bf \hspace{3cm} JND Training \hspace{4cm} \bf Triplet Finetuning
%\includegraphics[width=0.42\textwidth]{pictures/main_1_1_1.png} &
%\includegraphics[width=0.26\textwidth]{pictures/main_2_2_2.png} &
%\includegraphics[width=0.24\textwidth]{pictures/main_3_3_3.png}\\

\vspace{-22ex}
%\begin{tabular}{|p{0.40\textwidth}|p{0.25\textwidth}|p{0.25\textwidth}|}
%\bf \hspace{1cm} Self-supervised Audio encoder & \bf \hspace{3cm} JND Training & \hspace{1.5cm} \bf Triplet Finetuning
%\end{tabular}
\caption{\textbf{Metric Architecture}. We first train an audio encoder using contrastive learning, then train the loss-net on JND data~\cite{manocha2020differentiable}. Finally, we fine-tune the model on the newly collected dataset of triplet comparisons.}
\label{model_architecture}
\vspace{-2ex}
\end{figure*}
\fi

\vspace{-0.1in}
\section{Related Work}
\label{sec:format}

\subsection{Perceptual audio quality metrics}
\label{subsection:perceptualaudio}

\PESQ\,\cite{rix2001perceptual} and \VISQOL\,\cite{hines2015visqol} were some of the earliest models to approximate the human perception of audio quality.~Although useful, these had certain drawbacks: (i)~sensitivity to perceptually invariant transformations~\cite{hines2013robustness}; (ii)~narrow focus (\eg~telephony); and (iii)~non-differentiable, making it impossible to be optimized for as an objective in deep learning.~To overcome the last concern, researchers train differentiable models to approximate \PESQ~\cite{fu2019metricgan,zhang2018training}. The approach of Fu~\etal~\cite{fu2019metricgan} uses GANs to model \PESQ, whereas the approach of Zhang~\etal~\cite{zhang2018training} uses gradient approximations of \PESQ\ for training. Unfortunately, these approaches are not always optimizable, and may also fail to generalize to unseen perturbations.

Instead of using conventional metrics (\eg~\PESQ) as a proxy, Manocha~\etal~\cite{manocha2020differentiable} proposed \DPAM\ that was trained directly on a new dataset of human JND judgments. \DPAM\ correlates well with human judgment for small perturbations, but requires a large set of annotated judgments to generalize well across unseen perturbations.
Recently, Serra~\etal~\cite{serra2020sesqa} proposed \SESQA\ trained using the same JND dataset~\cite{manocha2020differentiable}, along with other datasets and objectives like \PESQ. Our work is concurrent with theirs, but direct comparison is not yet available.

%Researchers have trained differentiable neural network models to approximate \textit{PESQ} at each training iteration like the approach of Fu~\etal~\cite{fu2019learning} but it does not generalise to unseen audio perturbations. Other approaches like Generative adversarial networks (GANs) exist like MetricGAN~\cite{fu2019metricgan} but these are too task specific and may not always be optimizable.
%Manocha~\etal~\cite{manocha2020differentiable} proposed a deep perceptual audio metric (\DPAM) that was trained directly on human JND judgments. This metric correlates well with human judgment for small just-noticeable perturbations, but suffers from lack of generalization across speakers and content. 

\subsection{Representation learning}
\label{subsection:representation}
The ability to represent high-dimensional audio using compact representations has proven useful for many speech applications (\eg~speech recognition~\cite{conneau2020unsupervised}, audio retrieval~\cite{manocha2018content}).

\noindent {\bf Multi-dimensional learning:}
\label{subsection:mutidimlearning}
Learning representations to identify the underlying explanatory factors make it easier to extract useful information on downstream tasks~\cite{bengio2013representation}.
Recently in audio, Lee~\etal~\cite{lee2020disentangled} adapted Conditional Similarity Networks (CSN) for music similarity. The objective was to disentangle music into genre, mood, instrument, and tempo. Similarly, Hung~\etal~\cite{hung2018learning} explored disentangled representations for timbre and pitch of musical sounds useful for music editing. Chou~\etal~\cite{chou2018multi} explored the disentanglement of speaker characteristics from linguistic content in speech signals for voice conversion. Chen~\etal~\cite{chen2019audio} explored the idea of disentangling phonetic and speaker information for the task of audio representation.

%The main idea in CSN is to apply a masking function over the embedding space, where each mask corresponds to a different semantic dimension of similarity corresponding to musical notions such as genre, mood, instrument and tempo. Similarly, Hung et al.~\cite{hung2018learning} explored disentangled representations for timbre and pitch of musical sounds useful for music editing. ~\cite{huang2020unsupervised,chou2018multi}  explored disentanglement of speaker characteristics from the linguistic content in speech signal for voice conversion. Chen et al.~\cite{chen2019audio} explored the idea of disentangling phonetic and speaker information for the task of audio representation.

\noindent {\bf Contrastive learning}:
\label{subsection:contrastive}
%Contrastive learning is a self-supervised learning based approach which has recently got a lot of attention in computer vision~\cite{chen2020simple} after much success in natural language processing. 
Most of the work in audio has focused on audio representation learning~\cite{oord2018representation,chi2020audio}, speech recognition~\cite{schneider2019wav2vec}, and phoneme segmentation~\cite{kreuk2020self}. To the best of our knowledge, no prior method has used contrastive learning together with multi-dimensional representation learning to learn audio similarity.

\section{The CDPAM Metric}
\label{sec:model}
This section describes how we train the metric in three stages, 
depicted in Fig~\ref{model_architecture}: 
(a)~pre-train the audio encoder using contrastive learning;
(b)~train the loss-net on the perceptual JND data;
and 
(c)~fine-tune the loss-net on the new perceptual triplet data. 
%Our training model is depicted in Fig~\ref{model_architecture}.

\subsection{Dataset}
\label{subsection:dataset}

\noindent {\bf Self-supervised dataset:} We borrow ideas from the SimCLR framework~\cite{chen2020simple} for contrastive learning. SimCLR learns representations by maximizing agreement between differently augmented views of the same data example via a contrastive loss in the latent space. An audio is taken and transformations are applied to it to get a pair of augmented audio waveforms $x_i$ and $x_j$. Each waveform in that pair is passed through an encoder to get representations. These representations are further passed through a projection network to get final representations $z$.
%
%Then, a non-linear fully connected projection layer is applied to get representations $z$. 
The task is to maximize the similarity between these two representations $z_i$ and $z_j$ for the same audio, in contrast to the representation $z_k$ on an unrelated piece of audio $x_k$.

In order to combine multi-dimensional representation learning with contrastive learning, we force our audio encoder to output two sets of embeddings: \textit{acoustic} and \textit{content}. We learn these separately using contrastive learning.~To learn \textit{acoustic embedding}, we consider data augmentation that takes the same acoustic perturbation parameters but different audio content, whereas to learn \textit{content embedding}, we take different acoustic perturbation parameters but the same audio content. The pre-training dataset consists of roughly 100K examples.

\noindent {\bf JND dataset:}
We use the same dataset of crowd-sourced human perceptual judgments proposed by Manocha~\etal~\cite{manocha2020differentiable}. In short, the dataset consists of around 55K pairs of human subjective judgments, each pair coming from the same utterance, with annotations of whether the two recordings are \emph{exactly the same or different?}
Perturbations consist of additive linear background noise, reverberation, coding/compression, equalization, and various miscellaneous noises like pops and dropouts.

%For more information on the dataset and perturbations, refer to Manocha~\etal~\cite{manocha2020differentiable}.

\noindent {\bf Fine-tuning dataset:} To make the metric robust to large (beyond JND) perturbations, we create a triplet comparison dataset. 
%We use this dataset to fine-tune a model that was trained on the JND dataset.
%
This improves the generalization performance of the metric to include a broader range of perturbations and also enhances the ordering of the learned space.
%
%helps the metric to calibrate on longer range perturbations to better order the space. 
%
We follow the same framework and perturbations as Manocha~\etal~\cite{manocha2020differentiable}. This dataset consists of around 30K paired examples of crowd-sourced human judgments.

{
\begin{table*}
%\vspace{-1\baselineskip}
\setlength{\tabcolsep}{4pt}
\centering
% \resizebox{\textwidth}{!}{
 \begin{tabular}{l l c c c c c c c c}
 \toprule
 {\bf Type} & {\bf Name} & {\bf VoCo}~\cite{jin2017voco} & {\bf FFTnet}~\cite{jin2018fftnet}& {\bf BWE}~\cite{feng2019learning}& {\bf Dereverb} & {\bf HiFi-GAN} & {\bf PEASS} & {\bf VC} &{\bf Noizeus} \\
 \toprule
 
 \multirow{3}{*}{\bf Conventional}
  & {\bf MSE} & 0.18 & 0.18  & 0.00  & 0.14 & 0.00  & 0.25 &  0.00  & 0.20 \\
 & {\bf \PESQ} & 0.43 & 0.49  & 0.21  & 0.85 & \bf 0.70 & 0.71 & 0.56 & 0.68 \\
 & {\bf VISQOL} & 0.50 & 0.02 & 0.13 & 0.75 & 0.69 & 0.66 & 0.49 & 0.61 \\
 \cdashline{1-10}
 \multirow{1}{*}{\bf JND metric}
& {\bf \DPAM\ } & 0.71 & 0.63  & 0.61 & 0.45 & 0.30 & 0.63 & 0.45 & 0.10 \\
 \cdashline{1-10}
 
\multirow{1}{*}{\bf Ours (default)}

& {\bf \OURS\ } & \bf 0.73 & \bf 0.68  & \bf 0.65  & \bf 0.93 & 0.68  & \bf 0.74 & \bf 0.61 & \bf 0.71  \\ 
\bottomrule
 %\cdashline{2-10}
 %\multirow{2}{*}{\bf Ours (Ablation)}
 %& {\bf after JND} & 0.65 & 0.65  & \bf 0.79 & 0.74 & 0.42 & 0.71 & 0.61 (0.59) & 0.15 \\ 
 %& {\bf self-sup.} & 0.34 & 0.58  & 0.60   & 0.81 &0.46  & 0.73 & 0.50 (0.44) & 0.44 \\ 

 %\multirow{2}{*}{\bf Self-sup}
 % & {\bf VGGish} & 0.10 & 0.23  & -0.41 & -0.44  & 63.00 & 0.51 & 0.50 & 52.30 & 76.20 \\
 %& {\bf OpenL3} & 0.27 & 0.36 & 0.12 & 0.17 & 65.20 &  0.53 & 0.53 & 61.10 & 73.50  \\
 % \cdashline{1-14}
  
 %\bottomrule

\end{tabular}
% }
\vspace{-0.1in}
\caption{\textbf{Spearman correlation} of models with various MOS tests. Models include conventional metrics, DPAM and ours. Higher is better.}
\vspace{-2ex}
\label{table_mos}
\end{table*}
}

\vspace{-0.1in}
\subsection{Training and Architecture}
\vspace{-0.05in}
\noindent {\bf Encoder} Fig.~\ref{model_architecture}(a):
%Refer to Fig~\ref{fig1:subfig1}. 
The audio encoder consists of a 16 layer CNN with $15\times1$ kernels that is downsampled by half every fourth layer. We use global average pooling at the output to get a 1024-dimensional embedding, equally split into acoustic and content components.~The projection network is a small fully-connected network that takes in a feature embedding of size 512 and outputs an embedding of 256 dimensions.~The contrastive loss is taken over this projection. We use Adam~\cite{kingma2014adam} with a batch size of 16, learning rate of $10^{-4}$, and train for 250 epochs.

We use the \textit{NT-Xent loss} (Normalized Temperature-Scaled Cross-Entropy Loss) proposed by Chen~\etal~\cite{chen2020simple}. The key is to map the two augmented versions of an audio (positive pair) with high similarity and all other examples in the batch (negative pairs) with low similarity. For measuring similarity, we use cosine distance. For more information, refer to SimCLR~\cite{chen2020simple}.

\noindent {\bf Loss-network} Fig.~\ref{model_architecture}(b-c):
%Refer to Fig~\ref{fig1:subfig2} and~\ref{fig1:subfig3}. 
Our loss-net is a small 4 layer fully connected network that takes in the output of the audio encoder (\textit{acoustic embedding}) and outputs a distance (using the aggregated sum of the L1 distance of deep features). This network also has a small classification network at the end that maps this distance to a predicted human judgment. Our loss-net is trained using binary cross-entropy between the predicted value and ground truth human judgment.~We use Adam with a learning rate of $10^{-4}$, and train for 250 epochs. For fine-tuning on triplet data, we use \textit{MarginRankingLoss} with a margin of $0.1$, using Adam with a learning rate of $10^{-4}$ for 100 epochs.

As part of online data augmentation to make the model invariant to small delay, we decide randomly if we want to add a 0.25s silence to the audio at the beginning or the end and then present it to the network. This helps to provide shift-invariance property to the model, to disambiguate that in fact the audio is similar when time-shifted. To also encourage amplitude invariance, we also randomly apply a small gain (-20dB to 0dB) on the training data.

\iffalse
\begin{figure}[h!]
\centering
\includegraphics[width=1.08\columnwidth]{pictures/inclass_bwclass.png}
\vspace{-23ex}
\caption{Distribution Separation for in-class and bw-class groups}
\label{distribution_separation}
\end{figure}
\fi

\begin{table}[b!]
\centering
\setlength{\tabcolsep}{4pt}
\resizebox{\columnwidth}{!}{
 \begin{tabular}{l l c c c c}
 \toprule
 {\bf Type} & {\bf Name}  & {\bf FFTnet}& {\bf BWE}& {\bf HiFiGAN} & {\bf Simulated} \\
 \toprule
 %\cmidrule(lr){3-3} \cmidrule(lr){4-5} \cmidrule(lr){6-7} \cmidrule(lr){8-8} \cmidrule(lr){9-10} \cmidrule(lr){11-11} \cmidrule(lr){12-12} \cmidrule(lr){13-13} \cmidrule(lr){14-14}
 %& &\bf SC & \bf SC  & \bf 2AFC & \bf SC  & \bf 2AFC & \bf SC & \bf SC & \bf 2AFC & \bf SC & \bf SC & \bf SC & \bf 2AFC \\
 %\cmidrule(lr){1-14}
 
 \multirow{3}{*}{\bf Conventional}
  & {\bf MSE} & 55.0 & 49.0 & 70.2 & 43.0 \\
 & {\bf \PESQ} & 67.1 & 38.1 & 88.5 & 86.1 \\
 & {\bf \VISQOL} & 64.2 & 44.4 & 96.1 & 84.2 \\
 \cdashline{1-6}
 \multirow{1}{*}{\bf JND metric}
& {\bf \DPAM\ } & 61.5 & \bf 87.7 & 93.2 & 71.8\\
 \cdashline{1-6}
 
\multirow{1}{*}{\bf Ours (default)}

& {\bf \OURS\ } & \bf 88.5 & 75.9 & \bf 96.5  & \bf 87.7 \\ 
 \bottomrule
 %\cdashline{2-6}
 %\multirow{2}{*}{\bf Ours (Ablation)}
 %& {\bf after JND} & 84.65 & 78.3 & \bf 96.8 & 81.8\\ 
 %& {\bf self-sup.} &  89.20 & 76.2 & \bf 97.6 & 68.6 \\ 

 %\multirow{2}{*}{\bf Self-sup}
 % & {\bf VGGish} & 0.10 & 0.23  & -0.41 & -0.44  & 63.00 & 0.51 & 0.50 & 52.30 & 76.20 \\
 %& {\bf OpenL3} & 0.27 & 0.36 & 0.12 & 0.17 & 65.20 &  0.53 & 0.53 & 61.10 & 73.50  \\
 % \cdashline{1-14}
  
 %\bottomrule

\end{tabular}
}
%\vspace{-0.1in}
\caption{\textbf{2AFC accuracy} of various models, including conventional metrics, DPAM and ours. Higher is better.}
%\vspace{-3ex}
\label{table_triplet}
\end{table}

\section{Experiments}
\label{sec:experiments}
\iffalse
To show that our metric is robust and generalizes well to unseen perturbations, we show: i)~correlation to subjective ratings across 9 publicly available datasets; ii)~robustness to audio content variations and monotonocity with increasing noise; iii)~improvements to existing state of the art synthesis and denoising systems using our loss.
%To show that our metric is able to overcome the limitations of \textit{DPAM}, we show: 1) Robustness to content variation; 2) Correlation to subjective ratings across 9 publicly available datasets; 3) Improvements on existing state of the art vocoding and denoising systems using our loss.
\fi

\subsection{Subjective Validation}

We use previously published diverse third-party studies to verify that our trained metric correlates well with their task. We show the results of our model and compare it with \DPAM\ as well as more conventional objective metrics such as \textit{MSE}, \PESQ~\cite{rix2001perceptual}, and \VISQOL~\cite{hines2015visqol}.

We compute the correlation between the model's predicted distance with the publicly available MOS, using Spearman's Rank order correlation (SC). These correlation scores are evaluated per speaker where we average scores for each speaker for each condition.

As an extension, we also check for 2AFC accuracy where we present one reference recording and two test recordings and ask subjects \textit{which one sounds more similar to the reference?} Each triplet is evaluated by roughly 10 listeners. 2AFC checks for the exact ordering of similarity at per sample basis, whereas MOS checks for aggregated ordering, scale, and consistency. In addition to all evaluation datasets considered by Manocha~\etal~\cite{manocha2020differentiable}, we consider additional datasets:
% :(i) after self-supervised training; (ii) after JND training; and (iii) after triplet finetuning

\begin{enumerate}[leftmargin=0.33cm]
   \item \textbf{Dereverberation}~\cite{su2019perceptually}:
   consists of MOS tests to assess the performance of 5 deep learning-based speech enhancement methods.
   
   \item \textbf{HiFi-GAN}~\cite{su2020hifi}:
   consists of MOS and 2AFC scores to assess improvement across 10 deep learning based speech enhancement models (denoising and dereverberation).
   \item \textbf{PEASS}~\cite{emiya2011subjective}: consists of MOS scores to assess audio source separation performance across 4 metrics: \textit{global quality}, \textit{preservation of target source}, \textit{suppression of other sources}, and \textit{absence of additional artifacts}. Here, we only look at \textit{global quality}.
   
   \item \textbf{Voice Conversion (VC)}~\cite{lorenzo2018voice}: consists of tasks to judge the performance of various voice conversion systems trained using parallel (\textit{HUB}) and non-parallel data (\textit{SPO}). Here we only consider \textit{HUB}.
   
   \item \textbf{Noizeus}~\cite{hu2007subjective}: consists of a large scale MOS study of non-deep learning-based speech enhancement systems across 3 metrics: \textit{SIG}-speech signal alone; \textit{BAK}-background noise; and \textit{OVRL}-overall quality. Here, we only look at \textit{OVRL}.
  
\end{enumerate}

\noindent
Results are displayed in Tables~\ref{table_mos} and~\ref{table_triplet}, in which our proposed metric has the best performance overall. Next, we summarize with a few observations:

\begin{itemize} [leftmargin=0.33cm]
    
    \item Similar to findings by Manocha~\etal~\cite{manocha2020differentiable}, conventional metrics like \PESQ\ and \VISQOL\ perform better on measuring large distances (\eg~\textit{Dereverb}, \textit{HiFi-GAN}) than subtle differences (\eg~\textit{BWE}), suggesting that these metrics do not correlate well with human perception when measuring subtle differences.
    
    \item 
    %model suffers from a naturaltension between cost of data acquisition and generalization 
    
    We observe a natural compromise between generalizing to large audio differences well beyond JND (\eg~FFTnet, VoCo, etc.) and focusing only on small differences (\eg~BWE). As we see, \OURS\ is able to correlate well across a wide variety of datasets, whereas \DPAM\ correlates best where the audio differences are near JND.
    \OURS\ scores higher than \DPAM\ on \textit{BWE} on MOS correlation, but has a lower 2AFC score suggesting that \DPAM\ might be better at ordering individual pairwise judgments closer to JND.
    
    %Slight tradeoff when generalizing to higher order perturbations or focussing on smaller perturbations.
    
    %\item As the metric is able to generalise across a wide variety of tasks ranging from SE to voice conversion from all possible types consisting of deeplearning based noises to conventional noises. We think that the perturbation set introduced by Manocha et al is variant and sufficient to generalise to real life noises.
    
    %\item As expected, training on the JND data leads the model to learn low-level JND-type distances, causing an increase in correlations with datasets having subtle differences (\eg~\textit{VoCo,BWE}) but decreasing correlations with datasets having large differences (\eg~\textit{Dereverb,HiFi-GAN,Noizeus}). Similarly, fine-tuning on triplet data leads the model to learn higher order distances, enforcing better ordering in space. We observe an increase in 2AFC scores for datasets having larger distances (\eg~\textit{Dereverb,Noizeus}) and a slight decrease in scores for datasets having small distances (\eg~\textit{BWE}). This finding also suggests that the perturbation set considered by Manocha et al.~\cite{manocha2020differentiable} is able to model a variety of perturbations.
    
    \item Compared to \DPAM, \OURS\ performs better across a wide variety of tasks and perturbations, showing higher generalizability across perturbations and downstream tasks.
    
\end{itemize}

\iffalse
\begin{table}[h!]
\centering
\setlength{\tabcolsep}{2pt}
\resizebox{1\columnwidth}{!}{
 \begin{tabular}{l c c c c c}
 \toprule
 {\bf Name}  & {\bf Com.Area$\downarrow$} & {\bf Mono.$\uparrow$} & {\bf VoCo$\uparrow$} & {\bf FFTnet$\uparrow$}& {\bf BWE$\uparrow$} \\
 \toprule
 %\cmidrule(lr){3-3} \cmidrule(lr){4-5} \cmidrule(lr){6-7} \cmidrule(lr){8-8} \cmidrule(lr){9-10} \cmidrule(lr){11-11} \cmidrule(lr){12-12} \cmidrule(lr){13-13} \cmidrule(lr){14-14}
 %& &\bf SC & \bf SC  & \bf 2AFC & \bf SC  & \bf 2AFC & \bf SC & \bf SC & \bf 2AFC & \bf SC & \bf SC & \bf SC & \bf 2AFC \\
 %\cmidrule(lr){1-14}
{\bf after finetune (default)} & \bf 0.32 & \bf 0.89 &  \bf 0.73 & \bf 0.68 & 0.71 \\ 
 %\bottomrule
 \cdashline{1-6}
  
 {\bf \hspace{3mm} after JND} & \bf 0.32 & 0.88 & 0.65 & 0.65 & \bf 0.79\\ 
 {\bf  \hspace{3mm} self-sup.} & 0.34 & 0.53 & 0.34 & 0.58 & 0.60\\

 \cdashline{1-6}
 {\bf  w/o mul-dim. rep.} & 0.38 & 0.88 & 0.62 & 0.17 & 0.61\\ 
 \cdashline{1-7}
 {\bf triplet mul-dim. rep.} & 0.52 & 0.72 & 0.37 & 0.10 & 0.40\\ 
 \hline
 \hline
 {\bf \DPAM\ (prev. metric)} & 0.76 & \bf 0.89 &  - & - & -   \\ 
 \hline

 \bottomrule

\end{tabular}
}
\caption{\textbf{Ablation Studies}. ComArea refers to common area and Mono refers to monotonicity. Refer to Sec~\ref{subsection_ablation}}
\vspace{-2ex}
\label{table_ablation}
\end{table}
\fi

\begin{table}[b!]

\centering
\setlength{\tabcolsep}{2pt}
\resizebox{1\columnwidth}{!}{
 \begin{tabular}{l c c c c c}
 \toprule
 {\bf Name}  & {\bf ComArea$\downarrow$} & {\bf Mono$\uparrow$} & {\bf VoCo$\uparrow$} & {\bf FFTnet$\uparrow$}& {\bf BWE$\uparrow$} \\
 \toprule
 %\cmidrule(lr){3-3} \cmidrule(lr){4-5} \cmidrule(lr){6-7} \cmidrule(lr){8-8} \cmidrule(lr){9-10} \cmidrule(lr){11-11} \cmidrule(lr){12-12} \cmidrule(lr){13-13} \cmidrule(lr){14-14}
 %& &\bf SC & \bf SC  & \bf 2AFC & \bf SC  & \bf 2AFC & \bf SC & \bf SC & \bf 2AFC & \bf SC & \bf SC & \bf SC & \bf 2AFC \\
 %\cmidrule(lr){1-14}
 
  {\bf \DPAM~\cite{manocha2020differentiable}} & 0.76 & \bf 0.89 &  - & - & -   \\ 
\cdashline{1-6}
% \hline
% \hline
 
 {\bf self-sup. (triplet m.learning)} & 0.52 & 0.31 & 0.21 & 0.10 & 0.00\\ 
%  \cdashline{1-6}
 {\bf  contrastive w/o mul-dim. rep.} & 0.43 & 0.48 & 0.38 & 0.17 & 0.38\\ 
%  \cdashline{1-6}
 
 {\bf self-sup. (contrastive)} & 0.34 & 0.53 & 0.34 & 0.58 & 0.60\\ 
 {\bf \hspace{1mm} +JND} & \bf 0.32 & 0.88 & 0.65 & 0.65 & \bf 0.79\\ 
 {\bf \hspace{1mm} +JND+fine-tune (default)} & \bf 0.32 & \bf 0.89 &  \bf 0.73 & \bf 0.68 & 0.65 \\ 
 
  %\bottomrule
 %\cdashline{1-6}
  
 %{\bf \hspace{3mm} after JND} & \bf 0.32 & 0.88 & 0.65 & 0.65 & \bf 0.79\\ 

%  \hline
%  \hline
%   \hline

 \bottomrule

\end{tabular}
}
\caption{\textbf{Ablation studies}.~Sec~\ref{subsection_ablation} describes
common area, monotonicity, and Spearman correlations for 3 datasets.~$\uparrow$ or $\downarrow$ is better.}
%ComArea refers to common area and Mono refers to monotonicity. Refer to Sec~\ref{subsection_ablation}}
%\vspace{-4ex}
\label{table_ablation}
\end{table}

\subsection{Ablation study}
\label{subsection_ablation}

We perform ablation studies to better understand the influence of different components of our metric in Table~\ref{table_ablation}.~We compare our trained metric at various stages: (i)~after self-supervised training; (ii)~after JND training; and (iii)~after triplet finetuning (\textit{default}). 
To further compare amongst self-supervised approaches, we also show results of self-supervised metric learning using triplet loss.~To also show improvements due to learning multi-dimensional representations, we show results of a model trained using contrastive learning without \emph{content} dimension. 
The metrics are compared on (i)~robustness to content variations; (ii)~monotonic behavior with increasing perturbation levels; and (iii) correlation with subjective ratings from a subset of existing datasets.

%\begin{itemize} [leftmargin=0.33cm]
\noindent {\bf Robust to content variations}    
To evaluate robustness to content variations, we create a test dataset of two groups: one consisting of pairs of recordings that have the same acoustic perturbation levels but varying audio content; the other consisting of pairs of recordings having different perturbation levels and audio content. We calculate the common area between these normalized distributions. Our final metric has the lowest common area, suggesting that it is more robust to changing audio content. Decreasing common area also corresponds with increasing MOS correlations across downstream tasks, suggesting that the task of separating these two distribution groups may be a factor when learning acoustic audio similarity.

\noindent {\bf Clustered Acoustic Space:}     
To further quantify this learned space, we also calculate the Precision of Top $K$ retrievals which measures the quality of top $K$ items in the ranked list. Given 10 different acoustic perturbation groups - each group consisting of 100 files having the same acoustic perturbation levels but different audio content, we take randomly selected queries and calculate the number of correct class instances in the top $K$ retrievals. We report the mean of this metric over all queries ($MP^k$). \OURS\ gets $MP^{k=10}$ = 0.92 and $MP^{k=20}$ = 0.87, suggesting that these acoustic perturbation groups all cluster together in the learned space.

\noindent {\bf Monotonicity}
To show our metric's monotonicity with increasing levels of noise, we create a test dataset of recordings with different audio content and increasing noise perturbation levels (both \textit{individual} and \textit{combined} perturbations). We calculate SC between the distance from the metric and the perturbation levels. Both \DPAM\ and \OURS\ behave equally monotonically with increasing levels of noise. 

\noindent {\bf MOS Correlations}
Each of the key components of \OURS\, namely \emph{contrastive-learning}, \emph{multi-dimensional representation learning}, and \emph{triplet learning} have a significant impact on generalization across downstream tasks. 
%
%We also see significant improvements in MOS correlation due to multi-dimensional representation learning, making it a useful component of our metric.
%
Surprisingly, even our \textit{self-supervised} model has a non-trivial positive correlation with increasing noise as well as MOS correlation across datasets. This suggests that a self-supervised model not trained on any classification or perceptual task is still able to learn useful perceptual knowledge. This is true across datasets ranging from subtle to large differences suggesting that contrastive learning can be a useful pre-training strategy.

\iffalse
\vspace{-3ex}
\begin{figure}[h!]
\centering
\includegraphics[width=1\columnwidth]{pictures/vocoder.png}
\vspace{-4ex}
\caption{Subjective measures - MOS (left) and Pairwise Comparison (right) of vocoding methods. Our method is preferred over MelGAN~\cite{kumar2019melgan}. $\uparrow$ is better for ours.}
\label{pairwise-melgan}
\vspace{-5ex}
\end{figure}
\fi

\subsection{Waveform generation using our learned loss}

We show the utility of our trained metric as a loss function for the task of waveform generation.~We use the current state-of-the-art MelGAN~\cite{kumar2019melgan} vocoder. We train two models: i)~single speaker model trained on LJ~\cite{ito2017lj}, and ii)~cross-speaker model trained on a subset of VCTK~\cite{valentini2017noisy}.~Both the models were trained for around \emph{2 million} iterations until convergence.~We take the existing model and just add \OURS\ as an additional loss in the training objective.

%The only additional change we do is to add \OURS\ together with their GAN formulation.

\begin{figure}[h!]
\vspace{-1\baselineskip}
\includegraphics[width=\columnwidth]{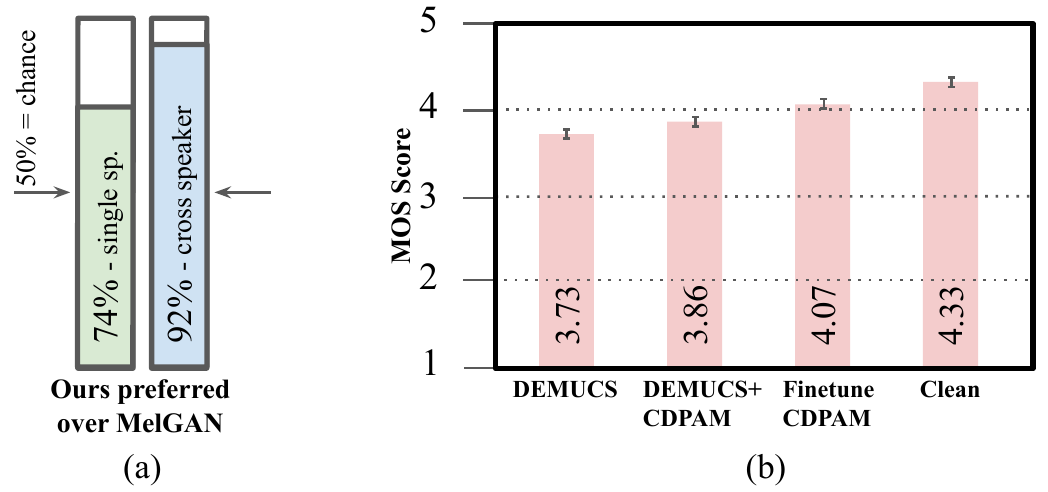}
\vspace{-0.25in}
\caption{\textbf{Subjective tests}: (a)~In pairwise tests, ours is typically preferred over MelGAN for single speaker and cross-speaker synthesis. (b)~MOS tests show denoising methods are improved by CDPAM.}
\vspace{-0.1in}
\label{se_vocoder}
\end{figure}

%We use the dataset available in~\cite{ito2017lj}, consisting of around 13,100 utterances of a single speaker for training. We use the same vocoder model as in~\cite{kumar2019melgan} for fair comparison. The only additional change we do is adding our loss function together with their GAN formulation. Hence, the new loss for Generator includes the multi-scale discriminator loss, feature matching loss and our perceptual loss.
We randomly select 500 unseen audio recordings to evaluate results. For the single-speaker model, we use LJ dataset, whereas for the cross-speaker model, we use the 20 speaker DAPS~\cite{mysore2014can} dataset.
%
%from single speaker LJ dataset and cross-speaker DAPS~\cite{mysore2014can} dataset and evaluate results. 
We perform A/B preference tests on Amazon Mechanical Turk (AMT), consisting of \emph{Ours vs baseline} pairwise comparisons. Each pair is rated by 6 different subjects and then majority voted to see which method performs better per utterance.~As shown in Fig~\ref{se_vocoder}(a), our models outperform the baseline in both categories. All results are statistically significant with p $<$ $10^{-4}$.~Our model is strongly preferred over the baseline, but the maximum improvement is observed in the cross-speaker scenario where our model performs best. Specifically, we observe that MelGAN enhanced with \OURS\ detects and follows the reference pitch better than the baseline.
%Issue of Melgan with multiple speakers, and how our model is able to ameliorate the issue. 

%We compare these new models with the original model on two metrics: quality of audio We randomly select 500 examples, and use those to judge the quality of the models. We perform objective and subjective tests to judge improvement. Figure x shows the \textit{PESQ vs Epoch} plot. 

%We also perform A/B preference tests on AMT, consisting of \textit{Ours vs baseline} pairwise comparisons. Each pair is rated by 15 different turkers and then majority voted to see which method performs better. Results are shown in Figure x.  All results are statistically significant with x.

\begin{table}[h!]

\resizebox{\columnwidth}{!}{ 
 \begin{tabular}{l c c c c c | c}
 \toprule
 {\bf } &  {\bf \PESQ} & {\bf STOI} & {\bf CSIG} & {\bf CBAK} & {\bf COVL} & {\bf MOS}
  \\ \toprule
  {\bf Noisy} 
 &  1.97 & 91.50 & 3.35 & 2.44 & 2.63 & 2.61 \\
 \midrule
  {\bf DEMUCS~\cite{defossez2020real}} 
 &  3.01 & 95.00 & 4.39 & 3.44 & 3.73 & 3.73\\
  {\bf DEMUCS+\OURS\ } 
 & \bf 3.10 & \bf 95.07 & \bf 4.46 & 3.55 & \bf 3.82 & 3.86 \\
 {\bf Finetune \OURS\ } 
 & 3.06 & 94.93 & 4.30 & \bf 3.56 & 3.70 & \bf 4.07\\

 \bottomrule
\end{tabular}
}
\vspace{-0.10in}
\caption{{\bf Evaluation of denoising} models using the VCTK~\cite{valentini2017noisy} test set with four objective measures and one subjective measure.}
\vspace{-2ex}
\label{table_denoising}
\end{table}

\iffalse

\begin{figure}[h!]
\vspace{-2\baselineskip}

\setlength{\w}{0.6\columnwidth}
\setlength{\tabcolsep}{2pt}
\begin{tabular}{c}

\subfloat[][\textit{Pairwise Comparison} of synthesis methods. Chance is 50\%(dotted line)]{\includegraphics[width=6cm, height=4cm,  keepaspectratio,center]{pictures/fig2/vocoder.png}
\label{fig2:subfig1}} \\
%\includegraphics[width=6cm, height=4cm,  keepaspectratio,center]{pictures/vocoder_pairwise.png}\\
%\bf a) \textit{Pairwise Comparison} of vocoding methods.\\\

\subfloat[][\textit{MOS} comparison of denoising methods]{\includegraphics[width=7cm, height=5cm,  keepaspectratio,left]{pictures/fig2/se_mos.png}
\label{fig2:subfig2}}

%\includegraphics[width=7cm, height=5cm,  keepaspectratio,left]{pictures/se_mos.png}\\
%\bf b) \textit{MOS} of denoising methods.

\end{tabular}
\vspace{-2ex}
\caption{\textbf{Subjective listening tests}: Our method is preferred over the baseline. $\uparrow$ is better for ours.}
\label{subjective_table}
\vspace{-4ex}
\end{figure}
\fi

\subsection{Speech Enhancement using our learned loss}

To further demonstrate the effectiveness of our metric, we use the current state-of-the-art DEMUCS architecture based speech denoiser~\cite{defossez2020real} and supplement \OURS\ in two ways: (i)~\textit{DEMUCS+CDPAM}: train from scratch using a combination of L1, multi-resolution STFT and \OURS; and (ii)~\textit{Fintune \OURS}: pre-train on L1 and multi-resolution STFT loss and finetune on \OURS. The training dataset consists of VCTK~\cite{valentini2017noisy} and DNS~\cite{reddy2020interspeech} datasets. For a fair comparison, we only compare real-time (causal) models.

We randomly select 500 audio clips from the VCTK test set and evaluate scores on that dataset. We evaluate the quality of enhanced speech using both objective and subjective measures. For the objective measures, we use: i) \PESQ\ (from 0.5 to 4.5); (ii) Short-Time Objective Intelligibility (\textit{STOI}) (from 0 to 100); (iii) \textit{CSIG}: MOS prediction of the signal distortion attending only to the speech signal (from 1 to 5); (iv) \textit{CBAK}: MOS prediction of the intrusiveness of background noise (from 1 to 5); (v) \textit{COVL}: MOS prediction of the overall effect (from 1 to 5). We compare the baseline model with both our models. Results are shown in Table~\ref{table_denoising}.

For subjective studies, we conducted a MOS listening study on AMT where each subject is asked to rate the sound quality of an audio snippet on a scale of 1 to 5, with \emph{1=Bad, 5=Excellent}. In total, we collect around 1200 ratings for each method. We provide studio-quality audio as reference for high-quality, and the input noisy audio as low-anchor.
%
%Each pair is rated by x different turkers and then majority voted to see which method performs better.
As shown in Fig~\ref{se_vocoder}(b), both our models perform better than the baseline approach. We observe that our \textit{Finetune} \OURS\ model scores the highest MOS score.
This highlights the usefulness of using \OURS\ in audio similarity tasks. Specifically, \OURS\ can identify and eliminate minor human perceptible artifacts that are not captured by traditional losses.~We also note that higher objective scores do not guarantee higher MOS, further motivating the need for better objective metrics.
%We also note that a higher PESQ score doesn't guarantee higher MOS, further motivating the need for better objective metrics.
%The baseline causal model has an aliasing artifact that our model is able to get rid of, which is why we observe higher MOS scores.

\section{Conclusion and Future work}

In this paper, we present \OURS, a contrastive learning-based deep perceptual audio metric that correlates well with human subjective ratings across tasks.~The approach relies on multi-dimensional and self-supervised learning to augment limited human-labeled data. We show the utility of the learned metric as an optimization objective for speech synthesis and enhancement, but it could be applied in many other applications.
We would like to extend this metric to include content similarity as well, in general going beyond acoustic similarity for applications like music similarity. Though we showed two applications of the metric, future works could also explore other applications like audio retrieval and speech recognition.

% -------------------------------------------------------------------------
%\vfill
%\ninept
\clearpage
\pagebreak
\bibliographystyle{IEEEbib}

\bibliography{mybib}

\end{document}